\documentclass[pra,twocolumn,showpacs,floatfix,superscriptaddress,aps]{revtex4}
\usepackage{amsmath}
\usepackage{makeidx}
\usepackage{amssymb}
\usepackage{graphicx}

\usepackage[usenames]{color}
\usepackage[normalem]{ulem}

\newcommand{\beq}{\begin{equation}}
\newcommand{\eeq}{\end{equation}} 
\newcommand{\beqa}{\begin{eqnarray}}
\newcommand{\eeqa}{\end{eqnarray}} 
\begin{document}


\title{Bright dipolar Bose-Einstein condensate
 soliton  mobile in a direction perpendicular to polarization}

\author{ S. K. Adhikari 
} 
\affiliation{
Instituto de F\'{\i}sica Te\'orica, UNESP - Universidade Estadual Paulista, 01.140-070 S\~ao Paulo, S\~ao Paulo, Brazil
} 

\begin{abstract}

We demonstrate  stable, robust,  bright 
dipolar Bose-Einstein condensate (BEC) solitons,   moving in a direction perpendicular to the 
polarization direction,  
formed due to  
dipolar interaction  for repulsive  
contact interaction. At medium  velocity the head on collision of two such solitons is found to be quasi  elastic with practically no deformation.  Upon small perturbation the solitons are found to exhibit sustained breathing oscillation. 
 The findings are illustrated by numerical simulation 
{using the 3D mean-field Gross-Pitaevskii 
equation and a reduced 2D model }
in three and two
spatial  dimensions employing realistic   interaction parameters for a dipolar $^{164}$Dy BEC.

\end{abstract}


\maketitle

\section{Introduction}
 
A bright soliton is a self-bound object
that maintains its shape, while
traveling at a constant velocity in one dimension (1D), due to a cancellation of nonlinear attraction and dispersive
effects. 
  In our three-dimensional (3D) world only quasi-solitons are observed where a reduced (integrated) 1D
density exhibits soliton-like properties.
    Solitons   have been studied in Bose-Einstein condensates (BEC), water wave, nonlinear optics,  among others \cite{rmp}.
Experimentally, bright matter-wave solitons  were created in a BEC of
$^7$Li \cite{1}
 and
$^{85}$Rb atoms \cite{3} by turning the atomic
interaction attractive from repulsive using a Feshbach resonance   \cite{fesh}.


\begin{figure}[!b]

\begin{center}
\includegraphics[width=\linewidth,clip]{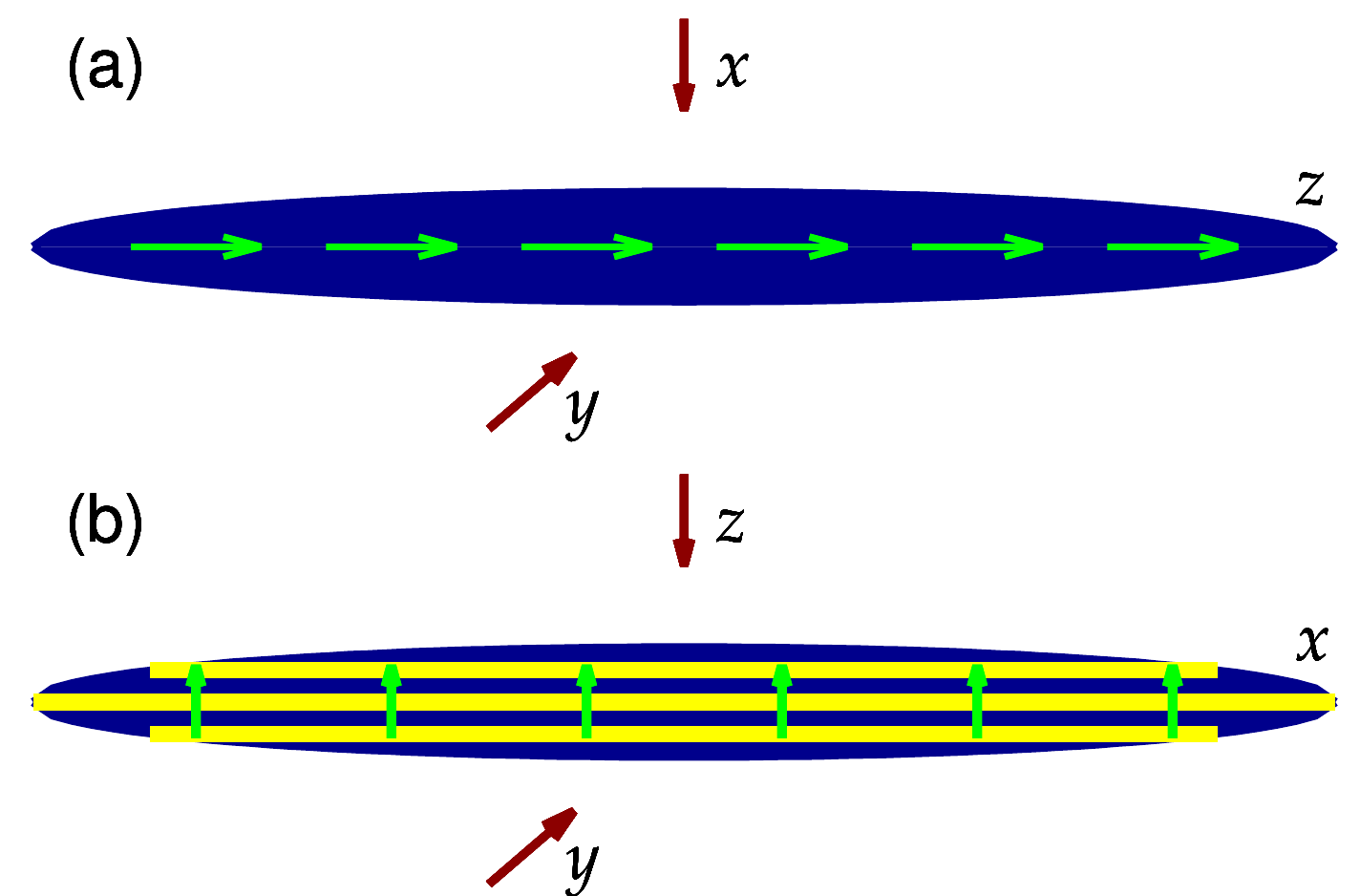}

\caption{ (Color online)  (a) The usual quasi-1D dipolar BEC soliton with polarization 
along the mobility direction $z$. (b) Proposed  dipolar BEC soliton with polarization 
direction $z$ perpendicular to the mobility direction $x$. The atoms in (b) are distributed in different regions 
perpendicular to the polarization direction $z$ shown by light (yellow) and dark (blue) stripes. 
The clear (green) arrows represent the polarization direction $z$ and the dark (brown) arrows represent the transverse trap directions. 
}\label{fig1} \end{center}

\end{figure}

{ 
The recent study  of  BECs of  $^{164}$Dy \cite{ExpDy,dy},   $^{168}$Er \cite{ExpEr}, and
  $^{52}$Cr \cite{cr,saddle} 
 atoms
with large magnetic dipole
moments has initiated  new investigations  of  BEC solitons in a different scenario.
One can  have dipolar BEC solitons for fully repulsive 
contact interaction \cite{1D}. {  Taking the polarization direction along the $z$ axis, quasi-1D solitons \cite{1D}   mobile along the $z$ axis have been established for a trap in the $x-y$ plane. 
Asymmetric  quasi-two-dimensional (quasi-2D) solitons \cite{2D} mobile in the $x-z$ plane with a trap along the $y$ axis have been confirmed. 
Quasi-2D 
vortex  solitons \cite{ol2D} 
have also been studied.
More recently, stable mobile  dark-in-bright  or  excited bright solitons have been demonstrated in 
quasi-1D \cite{dib1d} and quasi-2D \cite{dib2d} settings. In the former case the mobility direction is the $z$ axis with a trap in the $x-y$ plane and in the latter case the mobility direction  is the  $x-z$ plane with a trap along  the $y$ axis. }
 Dipolar BEC solitons can also be created in  periodic optical-lattice trap(s)
replacing the usual  
harmonic trap(s)
 in quasi-2D \cite{ol2D} and quasi-1D \cite{ol1D} set-ups.}

The nondipolar BEC solitons realized by a short-range attraction  can easily collapse thus 
making these solitons fragile.  
The dipolar BEC solitons stabilized  by a combination of long-range 
dipolar attraction stopping the atoms to escape and short-range repulsion inhibiting the collapse
are robust.  The dipolar BEC solitons thus can accommodate  a large number of atoms \cite{ol1D,ol2D,1D}. 
   The dipolar  interaction  is attractive along the polarization direction 
and repulsive in the transverse plane. Hence a quasi-1D dipolar BEC soliton free to move along the polarization direction $z$ naturally results \cite{1D} under a confining trap in the transverse plane.   However, the   dipolar interaction is not entirely repulsive in a plane perpendicular to the 
polarization direction. If it were so, an extremely  disk-shaped dipolar BEC confined in a plane 
perpendicular to the polarization direction
would be unconditionally stable. Nevertheless,  such a disk-shaped dipolar BEC 
collapses beyond a critical number of atoms indicating an attractive dipolar interaction in this set up  \cite{jbohn}. 
Here we demonstrate 
that it is possible to have a quasi-1D  dipolar BEC soliton with a quasi-2D shape in the $x-z$ plane, free to move along the $x$ direction,
(in a plane perpendicular 
to the polarization direction $z$), and  trapped in the transverse plane ($y$ and $z$ directions).

An usual quasi-1D dipolar BEC soliton  with the mobility direction parallel to the polarization 
direction is shown in Fig. \ref{fig1} (a). The proposed soliton with mobility and polarization directions orthogonal to each other is shown in   Fig. \ref{fig1} (b). {The atoms in this soliton can be considered to be distributed in regions perpendicular to the polarization direction shown by dark and bright stripes
in Fig. \ref{fig1} (b) {  for a weaker trap along the $z$ direction.} Qualitatively speaking, the   dipolar interaction among atoms in the 
same stripe 
is  repulsive as parallel dipoles placed side by side in a plane perpendicular to the polarization direction repeal each other.  However, 
 the  dipolar interaction among atoms in different stripes 
has a dominant attractive part as 
 parallel dipoles placed head-to-tail in a linear chain attract each other.  Consequently, stable   solitons of Fig. \ref{fig1} (b) can be realized  when there is a net dipolar attraction. 
However, a dipolar BEC in all  shapes  
 collapses beyond a critical number of atoms \cite{jbohn,cr,1D,ol2D,ol1D}. 
Hence the present quasi-1D dipolar BEC soliton  with the mobility direction perpendicular  to the polarization direction  is possible for a number of atoms below a critical number.   }
  {  The trapping geometry in this study, for a soliton mobile along the $x$ direction and the trap in the $y-z$ plane,  is distinct from that in Refs. \cite{dib1d,1D} where the mobility direction is the $z$ axis with a trap in the $x-y$ plane. However, if the weak $z$ trap 
in removed in the present set up, the present trapping reduces to the quasi-2D traps of Refs. 
\cite{dib2d,2D}.}

The head-on collision between two such solitons is found to be
quasi  elastic at medium  velocities of few mm/s. 
In such a collision, two solitons  pass through each other without significant deformation.   { However,  as the velocity is further lowered, the collision becomes inelastic with visible deformation of the solitons during collision. 
The collision of solitons can be completely elastic only in 1D
integrable systems.}  The solitons are found to exhibit sustained breathing oscillation upon 
small perturbation confirming their stability.

{ In Sec. II the time-dependent 3D mean-field model
for the dipolar BEC soliton is presented.
{ In the presence of a strong trap in the 
$y$ direction, 
a 2D reduction of this model 
appropriate for  the $x-z$ plane is also derived.   }
 The
results of numerical calculation are exhibited in Sec. III.
     Finally, in Sec. IV a
brief summary of our findings is presented.
}

\section{Mean-field model}  
   \label{II}

At ultra-low temperatures the properties of a dipolar condensate of $N$ 
atoms, each of mass $m$, can be described by the mean-field 
GP equation with nonlocal nonlinearity of the 
form:~\cite{cr,Santos01}
\begin{align}
i\hbar\frac{\partial \phi({\mathbf r},t)}{\partial t} &= 
\big[-\frac{\hbar^2}{2m}\nabla^2+V_{\text{trap}}({\mathbf r}) + \frac{4\pi\hbar^2a N}{m}| 
\phi({\mathbf r},t)|^2 \nonumber \\& + N \int U_{\mathrm{dd}}({\mathbf  r}-{\mathbf r}') 
\left\vert\phi({\mathbf r}',t)\right\vert^2 d{\mathbf r}' 
\big]\phi({\mathbf r},t),
\label{eqn:dgpe}
\end{align}
where $\int d{\bf r}  \vert  \phi({\mathbf r},t) \vert  ^2=1.$ 
 The trapping potential, $V_{\text{trap}}$ is 
assumed to be fully  asymmetric of the form
\begin{align}\label{trap}
V_{\text{trap}}({\mathbf r}) = \frac{1}{2} m\omega^2\left(\gamma^2 x^2+
\nu^2 y^2+ \alpha^2 z^2 \right) 
\end{align}
where   
$\omega$ is  the trap frequency and $\gamma$, $\nu$ and $\alpha$ are anisotropy parameters,  $a$ the atomic scattering length.  
The dipolar interaction, for magnetic dipoles, is \cite{cr} 
\begin{align}
U_{\mathrm{dd}}(\bf R)\equiv \frac{\mu_0 \mu^2}{4\pi}   V_{\mathrm{dd}}(\bf R) =\frac{\mu_0 \mu^2}{4\pi}\frac{1-3\cos^2 \theta}{ \vert  {\bf R} \vert  ^3},
\end{align}
where ${\bf R= r -r'}$ determines the relative position of dipoles and $\theta$ 
is the angle between ${\bf R}$ and the direction of polarization $z$, 
 $\mu_0$ is the permeability of free space 
and $ \mu$ is the dipole moment of an atom. The strength  of  dipolar interaction can be  expressed in terms  
 of a dipolar length  $a_{\mathrm {dd}}$ 
  defined  by  \cite{cr,saddle}
\begin{equation}
a_{\mathrm{dd}}\equiv\frac{ \mu_0  \mu^2  m}{12\pi \hbar^2}.
\end{equation}

For the formation of a   soliton, mobile  in the $x$ direction,  the parameters of the trap (\ref{trap})
are taken as $\gamma =0, $ and $\nu  > \alpha$. In the present study we take $\nu =1$.
A dimensionless GP equation for the dipolar BEC  soliton, mobile  in the $x$ direction,
can be written as   \cite{cr,1D}
\begin{align}& \,
 i \frac{\partial \phi({\bf r},t)}{\partial t}=
{\Big [}  -\frac{\nabla^2}{2 }
+
\frac{1}{2}(y^2+\alpha^2 z^2)
\nonumber \\  &  \,
+ g \vert \phi({\bf r},t) \vert^2
+ g_{\mathrm {dd}}
\int V_{\mathrm {dd}}({\mathbf R})\vert\phi({\mathbf r'},t)
\vert^2 d{\mathbf r}' 
{\Big ]}  \phi({\bf r},t),
\label{eq3}
\end{align}
where
$g=4\pi a N,$ 
$g_{\mathrm {dd}}= 3N a_{\mathrm {dd}},$ 
and where $\omega$  and $\omega\alpha$ are the frequency 
of the harmonic trap along $y$ and $z$ axes. 
In Eq. (\ref{eq3}), length is expressed in units of 
oscillator length  $l=\sqrt{\hbar/(m\omega)}$, 
energy in units of oscillator energy  $\hbar\omega$, probability density 
$|\phi|^2$ in units of $l^{-3}$, and time in units of $ 
t_0=1/\omega$.

For a strong trap in the $y$ direction and an useful 
quasi-2D mean-field model can be derived by integrating       out the $y$ dependence  
 assuming that the dynamics of the BEC in the $y$ direction
 is confined in the ground state \cite{luca}
\begin{align}
\phi_{1D}(y)=\frac{e^{-y^2 /2d_y^2}}{(\pi d_y^2)^{1/4}}, \quad d_y\equiv  \sqrt{\frac{1}
{\nu}}=1,
\end{align}
and we have for the wave function
\begin{align} \label{an2xz}
\phi({\bf r},t)\equiv \phi_{1D}(y) \times \phi_{2D}(\boldsymbol \rho{},t)
 ,
\end{align}
where   now $\boldsymbol \rho{}\equiv (x,z)$, and 
$\phi_{2D}(\boldsymbol \rho{},t)$ is the circularly-asymmetric 
effective 2D wave function for the 2D dynamics
and $d_y$ is the   harmonic oscillator length along the $y$ direction.
To derive the effective 2D equation for the disk-shaped dipolar BEC,
we use  ansatz (\ref{an2xz}) in Eq. (\ref{eq3}), multiply by the 
ground-state wave function $\phi_{1D}(y)$ and integrate over $y$ to get 
the 2D equation 
 \begin{align}\label{gpe2dxz}
&
i\frac{\partial \phi_{2D}(\boldsymbol \rho,t)}{\partial t}=\biggr[-\frac{\nabla_\rho^2}{2}
+\frac{\alpha^2 z^2}{2}+\frac{4\pi aN \vert  \phi_{2D} \vert  ^2}{\sqrt{2\pi}d_y}
\nonumber \\&
+ {3 a_{\mathrm{dd}}N}\int d \boldsymbol \rho' V_{\mathrm{dd}}^{2D}(\boldsymbol \rho -\boldsymbol \rho') |\phi_{2D}(\boldsymbol \rho{}', t)|^2
\biggr] \phi_{2D}(\boldsymbol \rho,t).
\end{align}
The dipolar interaction  $V_{\mathrm{dd}}^{2D}$
 is calculated in momentum space by the following convolution integral  \cite{ska,jb2}
\begin{align}
\int d \boldsymbol \rho' & V_{\mathrm{dd}}^{2D}(\boldsymbol \rho -\boldsymbol \rho') |\phi_{2D}(\boldsymbol \rho{}', t)|^2 \nonumber \\ &=
 \frac{4\pi}{3}\int \frac{d{\bf k}_\rho}{(2\pi)^2}
 e^{-i{\bf k}_\rho.\boldsymbol \rho}\widetilde n({\bf k}_\rho,t)j_{2D}
\biggr(\frac{k_\rho d_y}{\sqrt 2}\biggr),\\
\widetilde n({\bf k}_\rho,t)&=\int d\boldsymbol \rho e^{i{\bf k}_\rho.\boldsymbol \rho} \vert  \phi_{2D}(\boldsymbol \rho,t) \vert  ^2,\\
j_{2D}(\xi)&\equiv \frac{1}{2\pi}\int^\infty_{-\infty} dk_y \left[  \frac{3k_z^2}{{\bf k}^2}  -1 \right] |\widetilde n(k_y)|^2
\nonumber \\&
 =\frac{1}{\sqrt{2\pi}d_y}
[ -1+3\sqrt \pi \frac{k_z^2 d_y^2}{2\xi} e^{\xi^2}\{1-\text{erf}(\xi)\}],\\
\widetilde n(k_y)&=\int_{-\infty}^\infty dy e^{i k_y y}|\phi_{1D}(y)|^2= 
e^{-k_y^2d_y^2/4},
\end{align}
  where $k_\rho=\sqrt{k_z^2+k_x^2}$~ and $\xi={k_\rho d_y}/{\sqrt 2}$.
 
 \section{Numerical Results}
  We consider $^{164}$Dy
atoms in this study of  BEC solitons. 
  The magnetic moment of a $^{164}$Dy  
atom is  $ \mu_1 = 10\mu_B$
\cite{ExpDy} 
with 
$\mu_B$ the Bohr magneton leading to the dipolar lengths $a_{\mathrm {dd}}(^{164}$Dy$) \approx 132.7a_0$,    with $a_0$ the Bohr radius. 
The dipolar interaction in $^{164}$Dy
atoms is roughly   eight times larger than that in $^{52}$Cr
atoms with a dipolar length  $a_{\mathrm {dd}}\approx 15a_0$ 
\cite{cr}. We take $l\equiv \sqrt{\hbar/m\omega} =1$ $\mu$m. In a  $^{164}$Dy BEC this corresponds to an angular trap frequency $\omega =  2\pi \times 61.6$ Hz corresponding to $t_0= 2.6$ ms.

We solve the GP equations  (\ref{eq3}) and (\ref{gpe2dxz}) 
by the split-step 
Crank-Nicolson method
using both real- and imaginary-time propagation
  in  Cartesian coordinates  
using a space step of 0.1 $\sim$ 0.2
and a time step of 0.0004 $\sim$ 0.005 \cite{CPC,ska}.  The dipolar potential term is treated by Fourier transformation  in momentum space \cite{Santos01}.

\begin{figure}[!t]

\begin{center}
\includegraphics[width=.49\linewidth,clip]{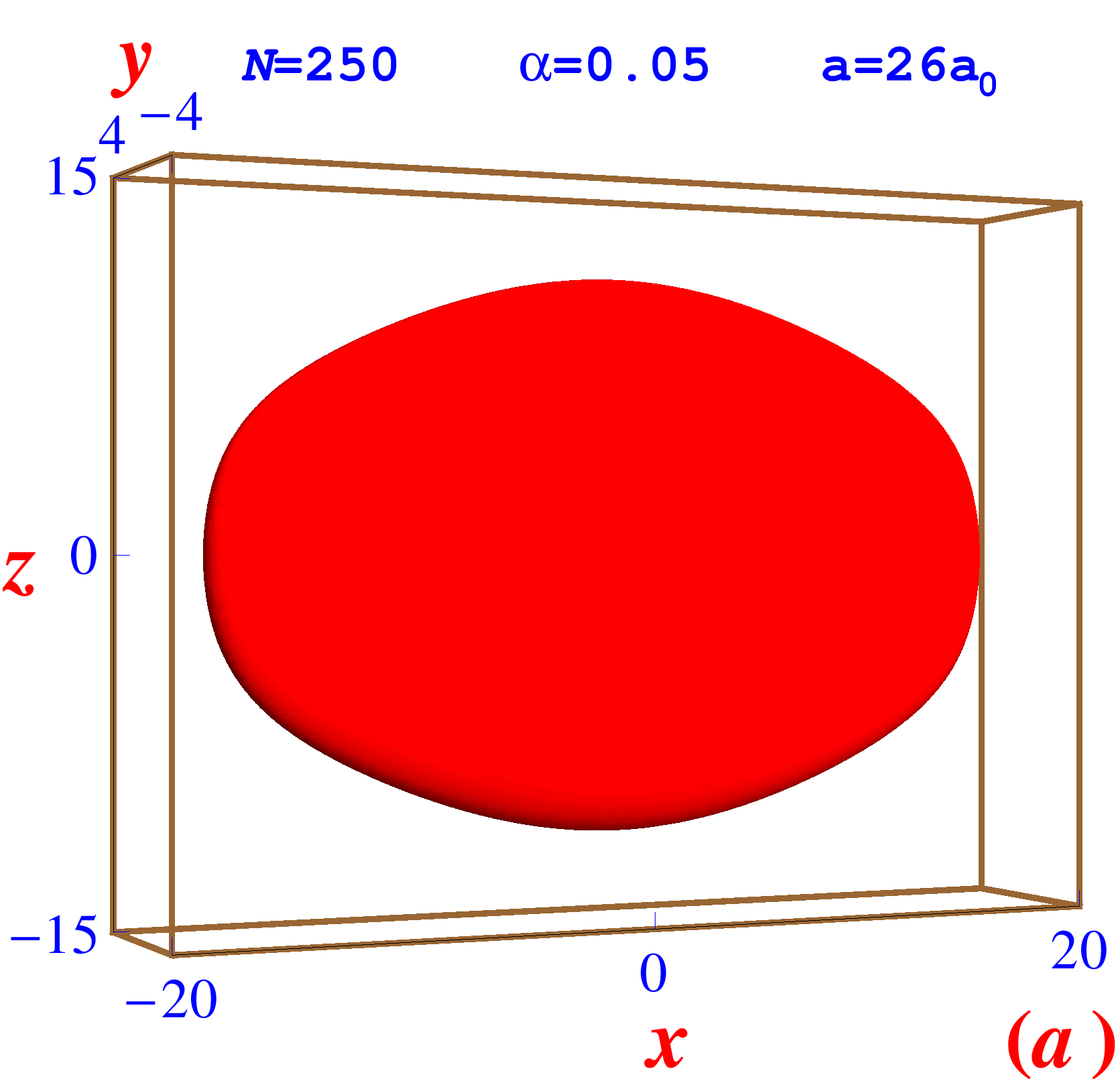}
\includegraphics[width=.49\linewidth,clip]{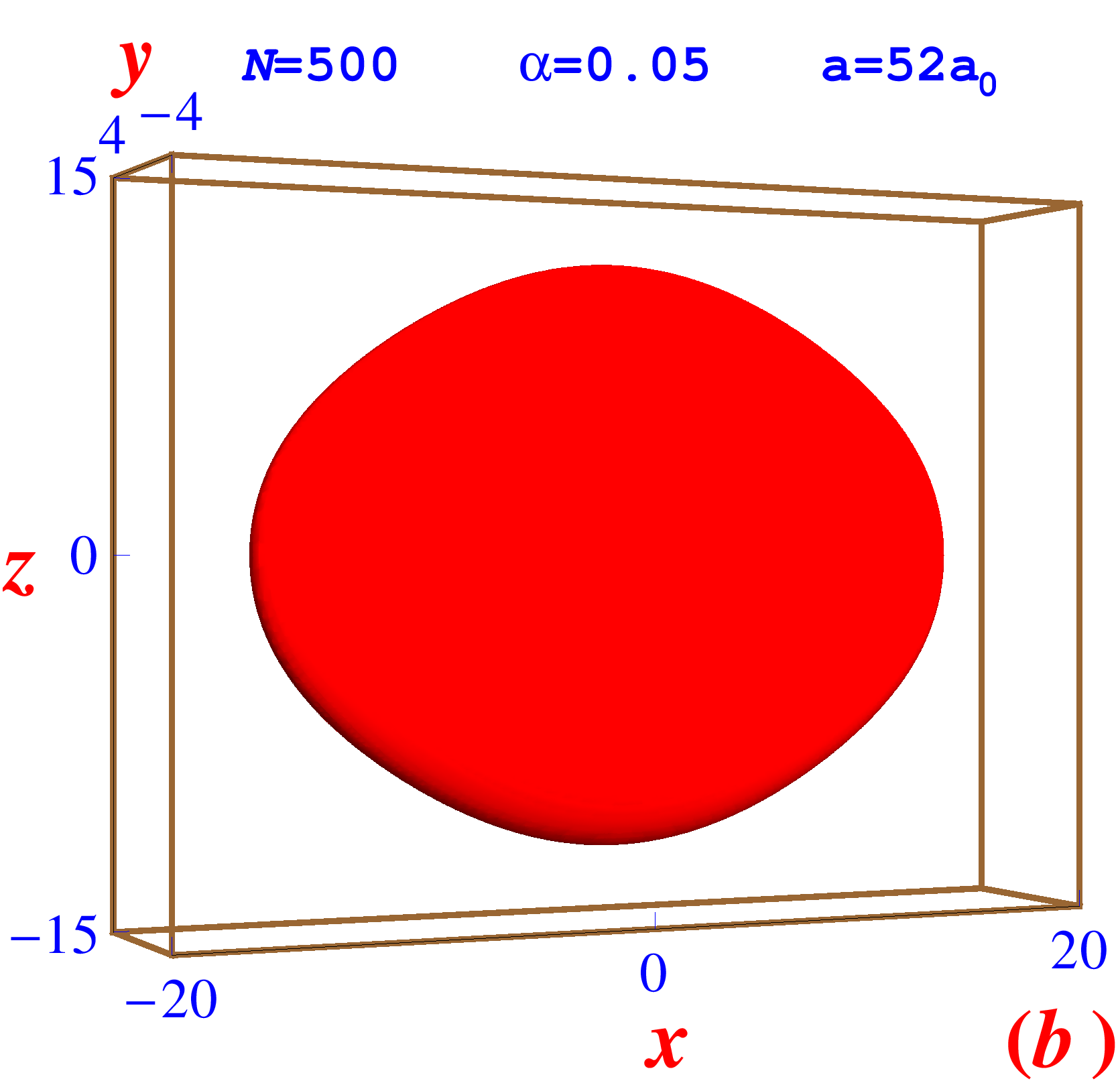}
\includegraphics[width=.49\linewidth,clip]{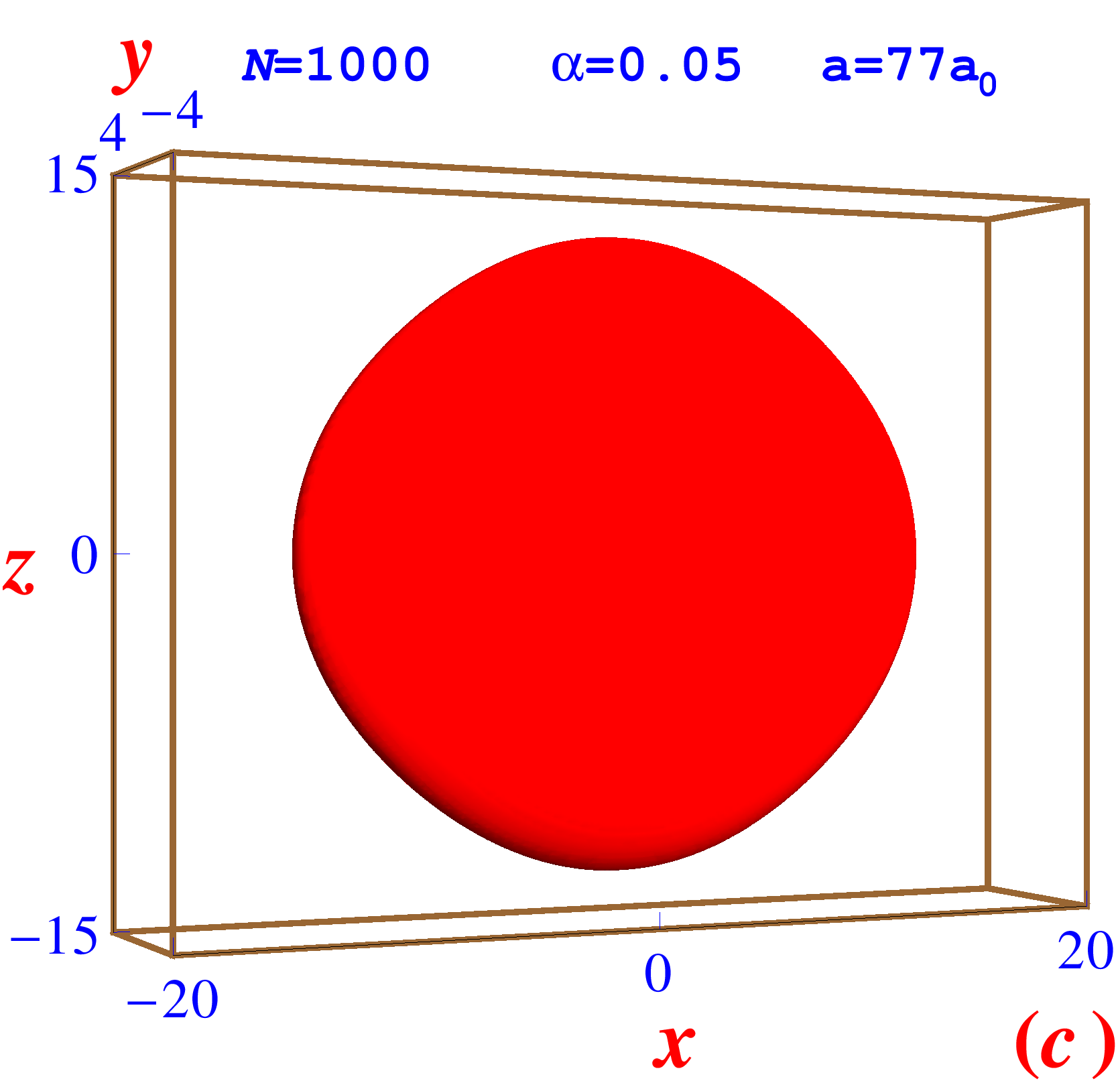}
\includegraphics[width=.49\linewidth,clip]{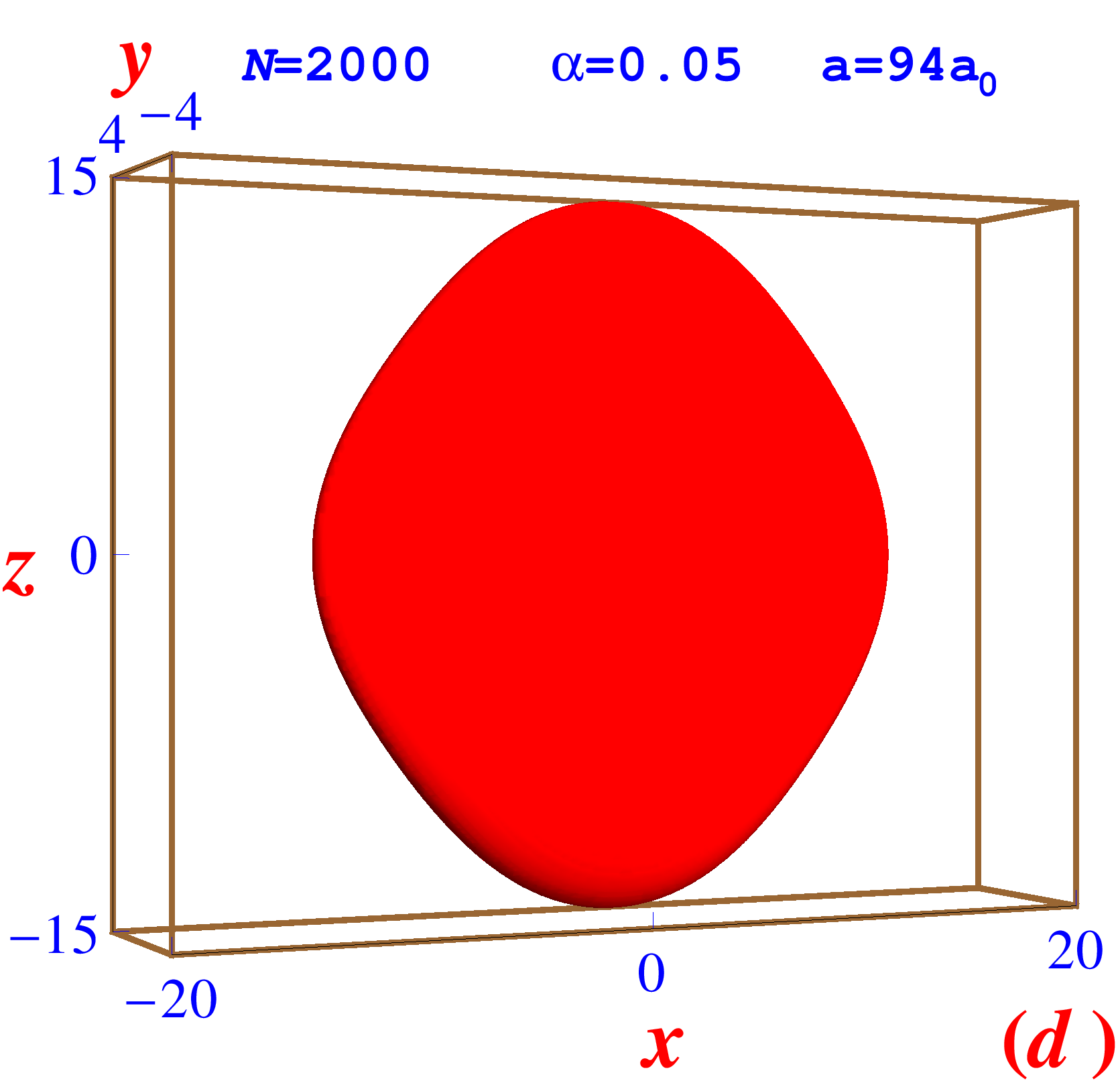}
\includegraphics[width=.7\linewidth,clip]{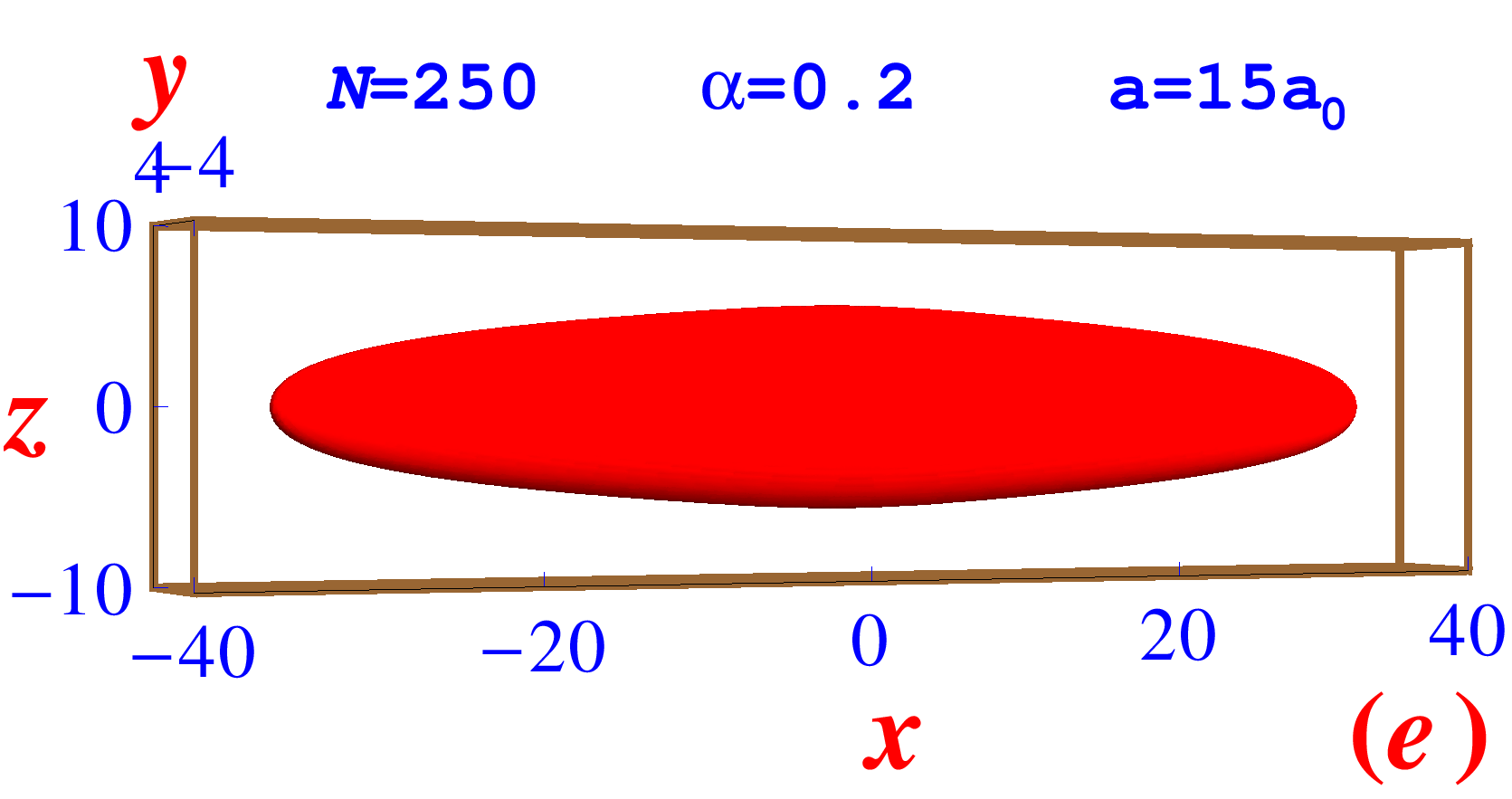}
\includegraphics[width=.7\linewidth,clip]{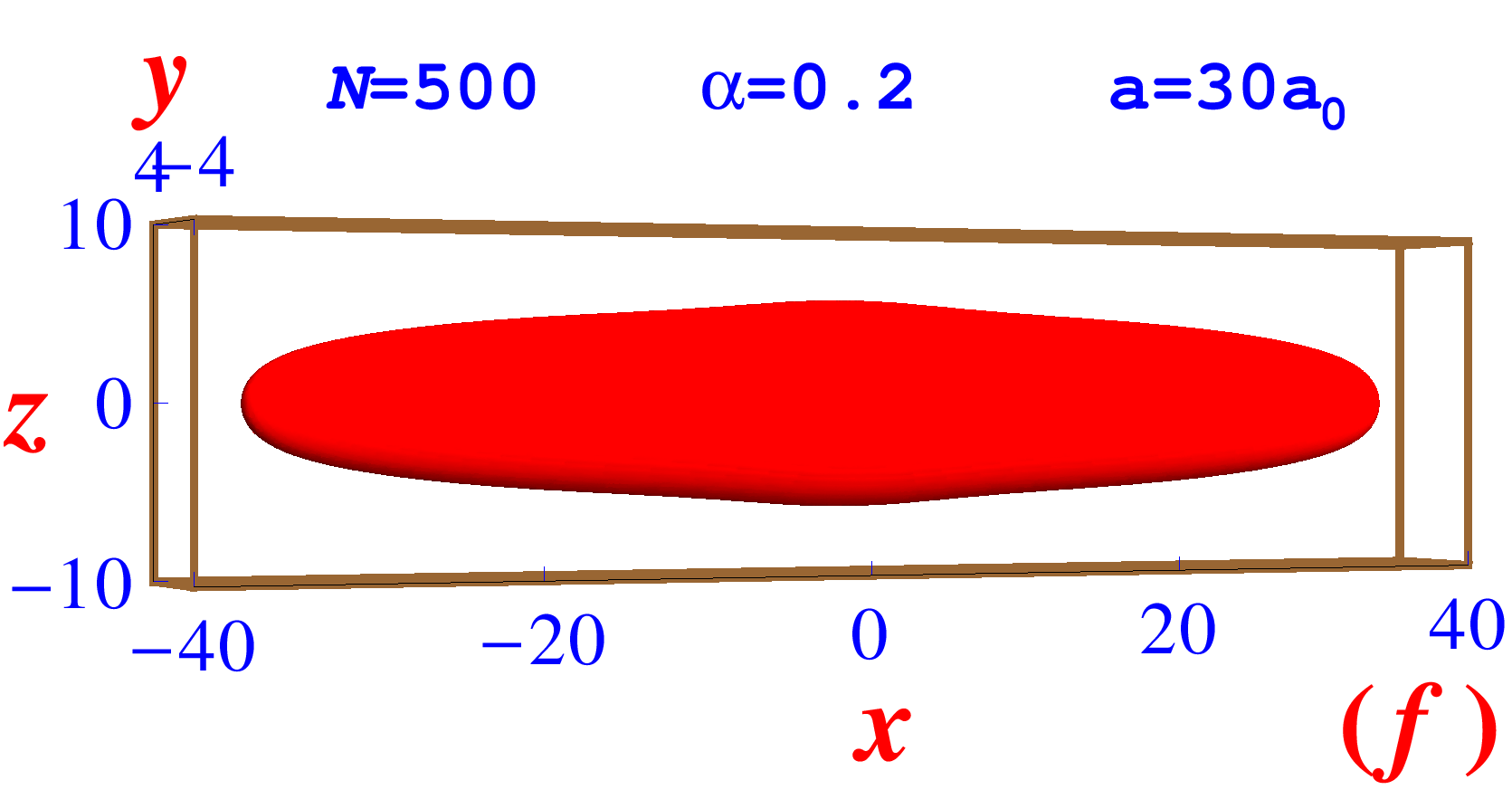}

\caption{ (Color online) 3D isodensity contour   $|\phi({\bf r})|^2$ of a bright soliton of  
$^{164}$Dy atoms for (a) $a= 26a_0, N =250, \alpha =0.05$,  (b)  $a= 52a_0, N= 500, \alpha=0.05$,  (c)  $a= 77a_0, N= 1000, \alpha =0.05$, and (d)  $a= 94a_0, N =1000, \alpha =0.05$,  (e)  $a= 15a_0, N =250, \alpha =0.2$, (f)  $a= 30a_0, N =500, \alpha =0.2$,
  The  lengths $x,y,$ and $z$ are in units of $l$ $(\equiv 1$ $\mu$m). The dimensionless
density on 
the contour is 0.00001. With present length scale $l$ 
this density  corresponds to  $ 10^{7}$ atoms/cm$^3$. }

\label{fig2} \end{center}

\end{figure}


The 3D GP equation (\ref{eq3}) is solved for different values of the number of atoms $N$,
scattering length $a$, and dipolar 
length $a_{\mathrm{dd}}$. We consider the trap anisotropy parameter {$\alpha =0.05$ and 0.2}  in this study. 
In Figs. \ref{fig2} (a) $-$ (f) we display the 3D isodensity contour of the present solitons for 
{differnet $N, a$ and $\alpha$.}  
A variation of the scattering length can be realized experimentally 
by  the Feshbach resonance technique \cite{fesh}.
 Because of the weak trap in the $z$ direction, the quasi-1D 
solitons have a quasi-2D shape in the $x-z$ plane, as can be seen in 
Fig. \ref{fig2}. { The quasi-2D shape tends to a quasi-1D one with the increase of 
trap anisotropy 
$\alpha$ as can be seen in Figs. \ref{fig2} (a) and (b). 
}

\begin{figure}[!t]

\begin{center}
\includegraphics[width=.49\linewidth,clip]{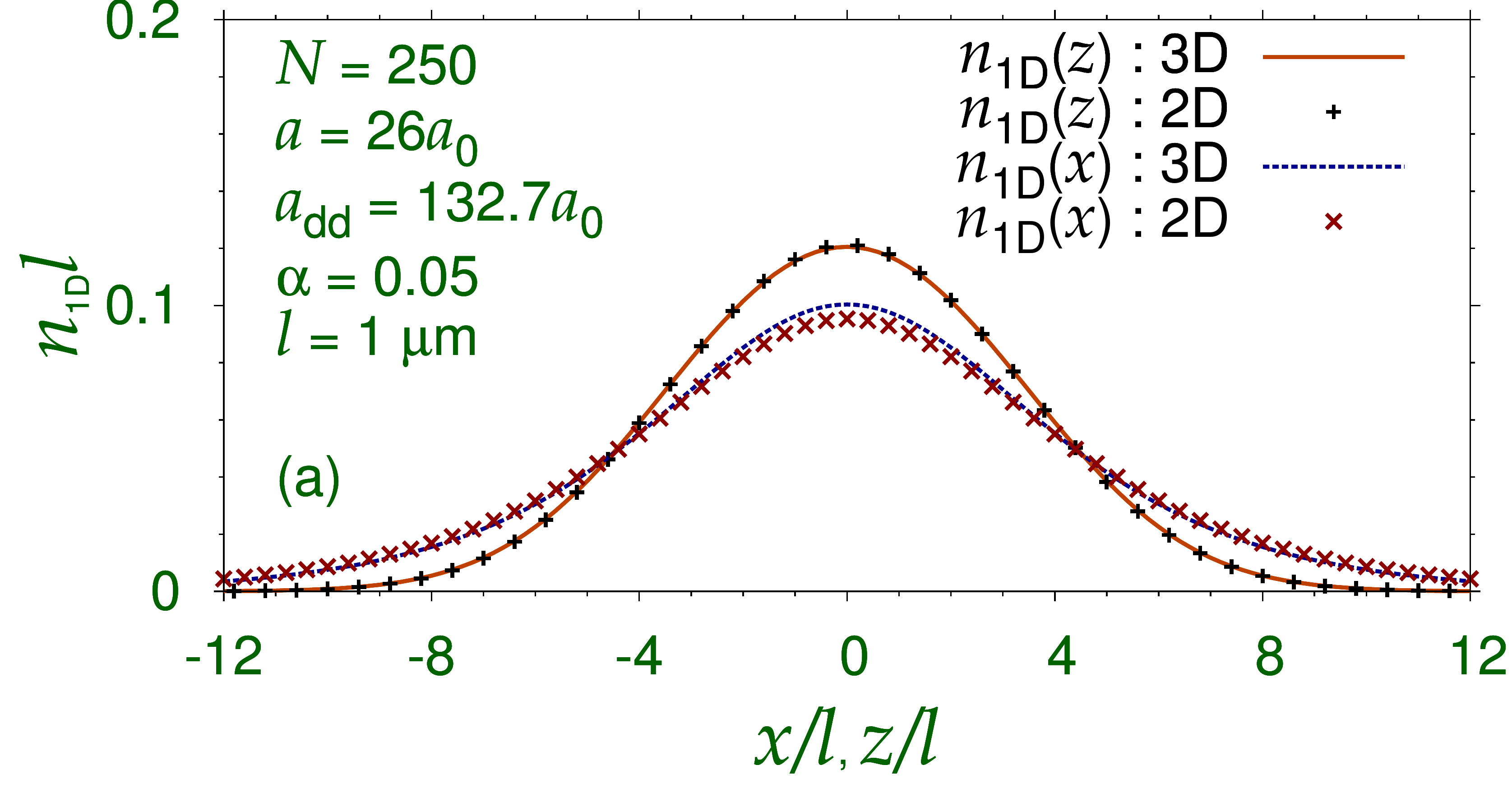}
\includegraphics[width=.49\linewidth,clip]{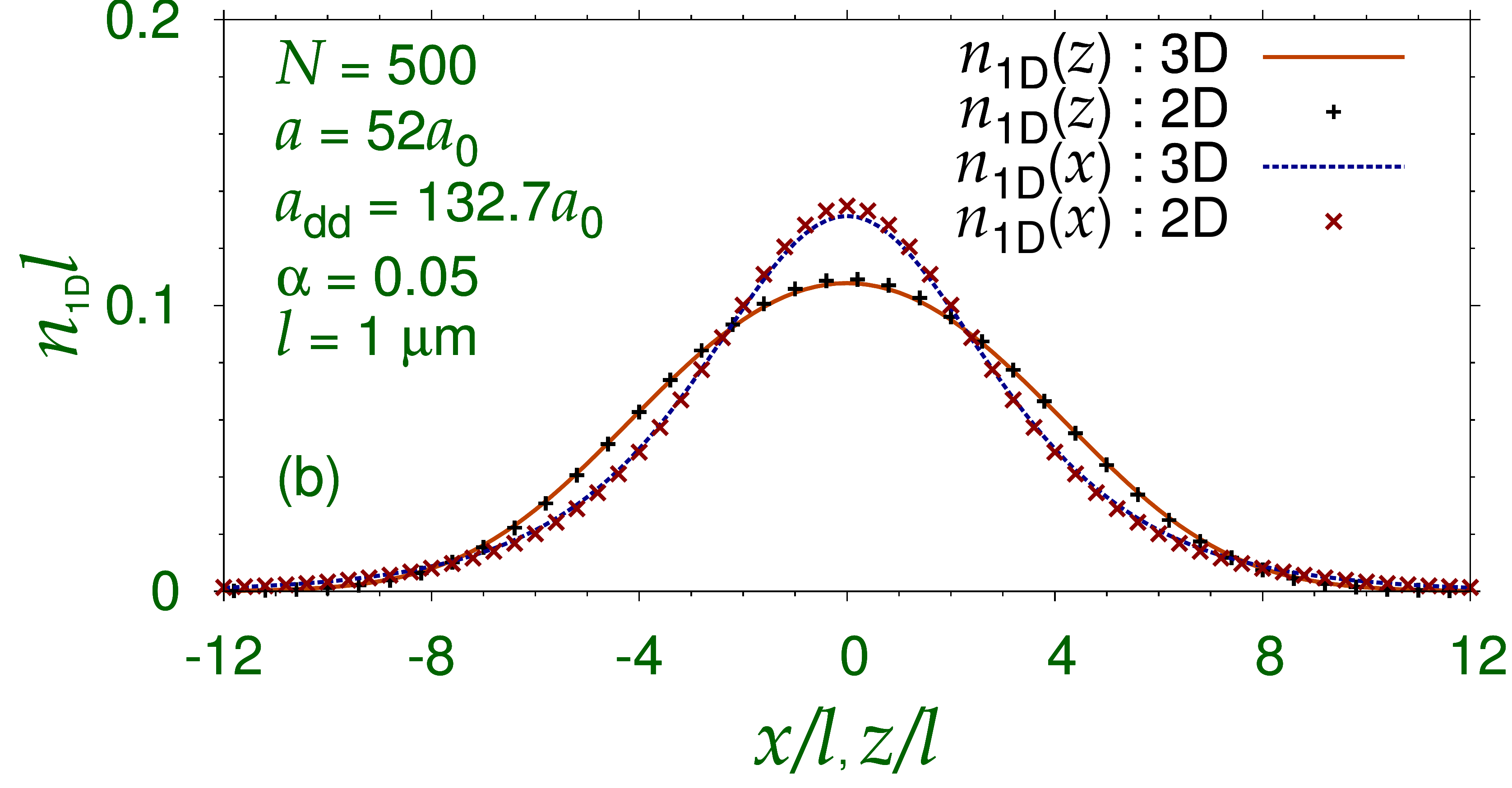}
\includegraphics[width=.49\linewidth,clip]{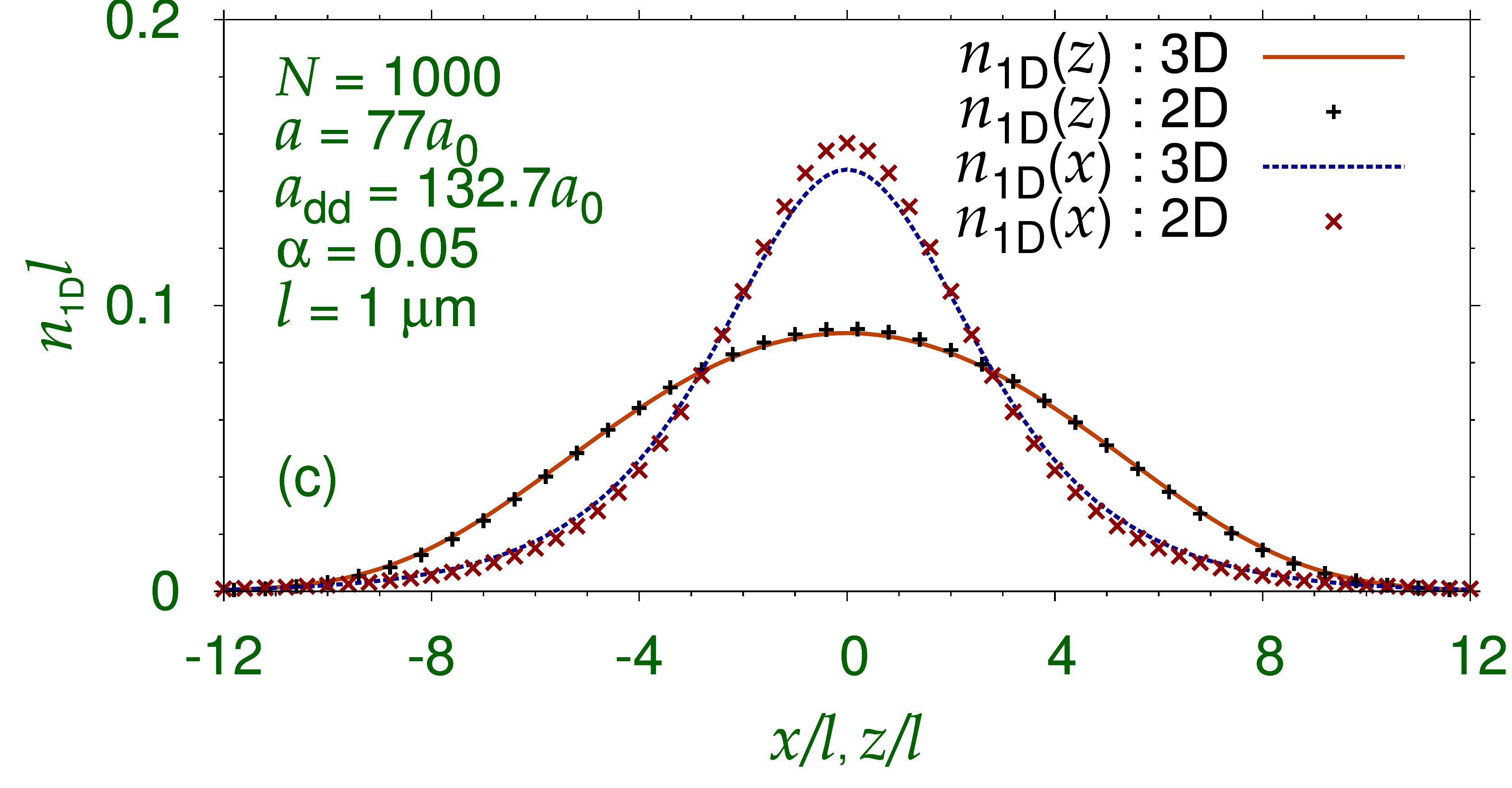}
\includegraphics[width=.49\linewidth,clip]{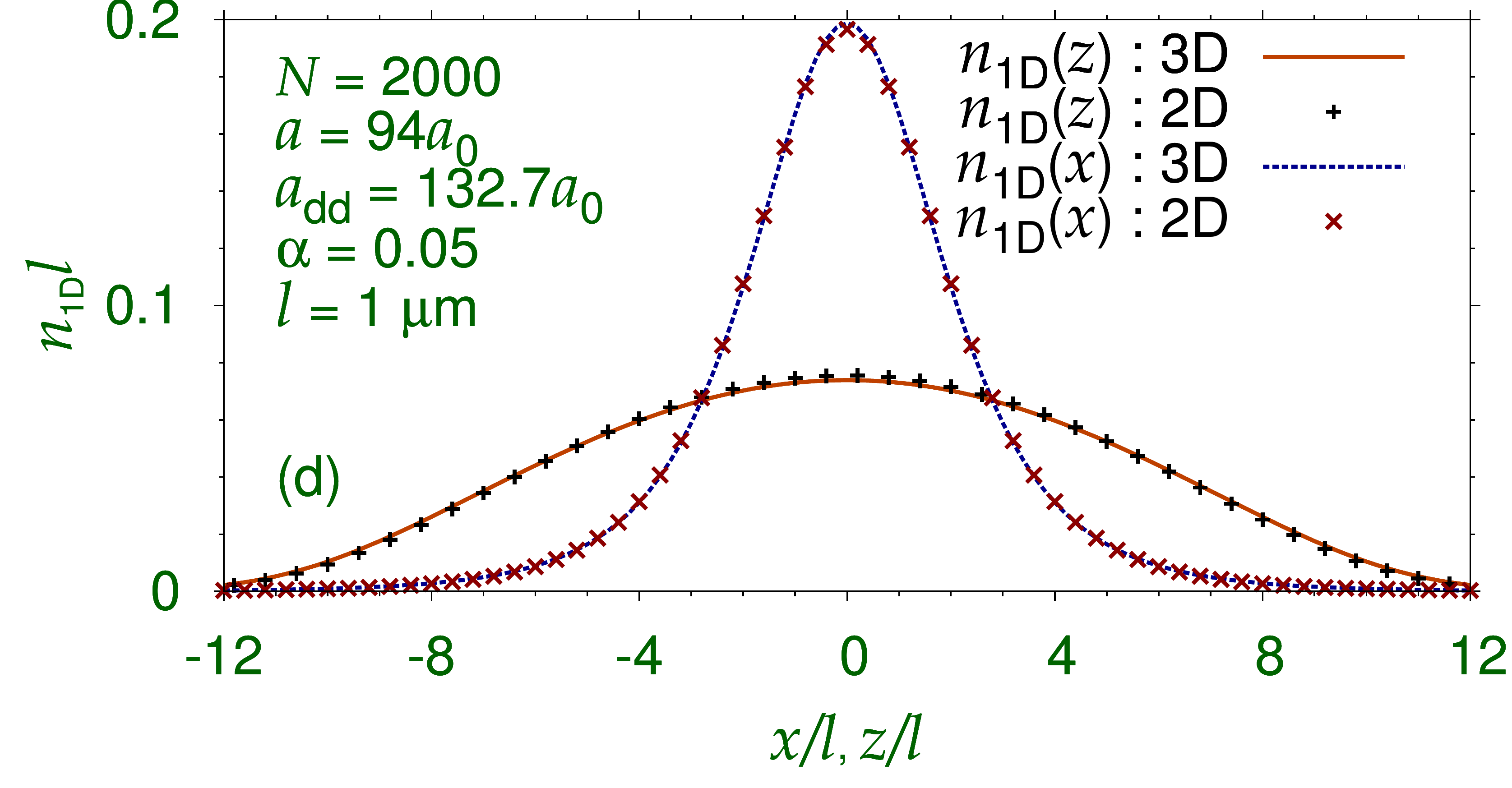}
\includegraphics[width=.49\linewidth,clip]{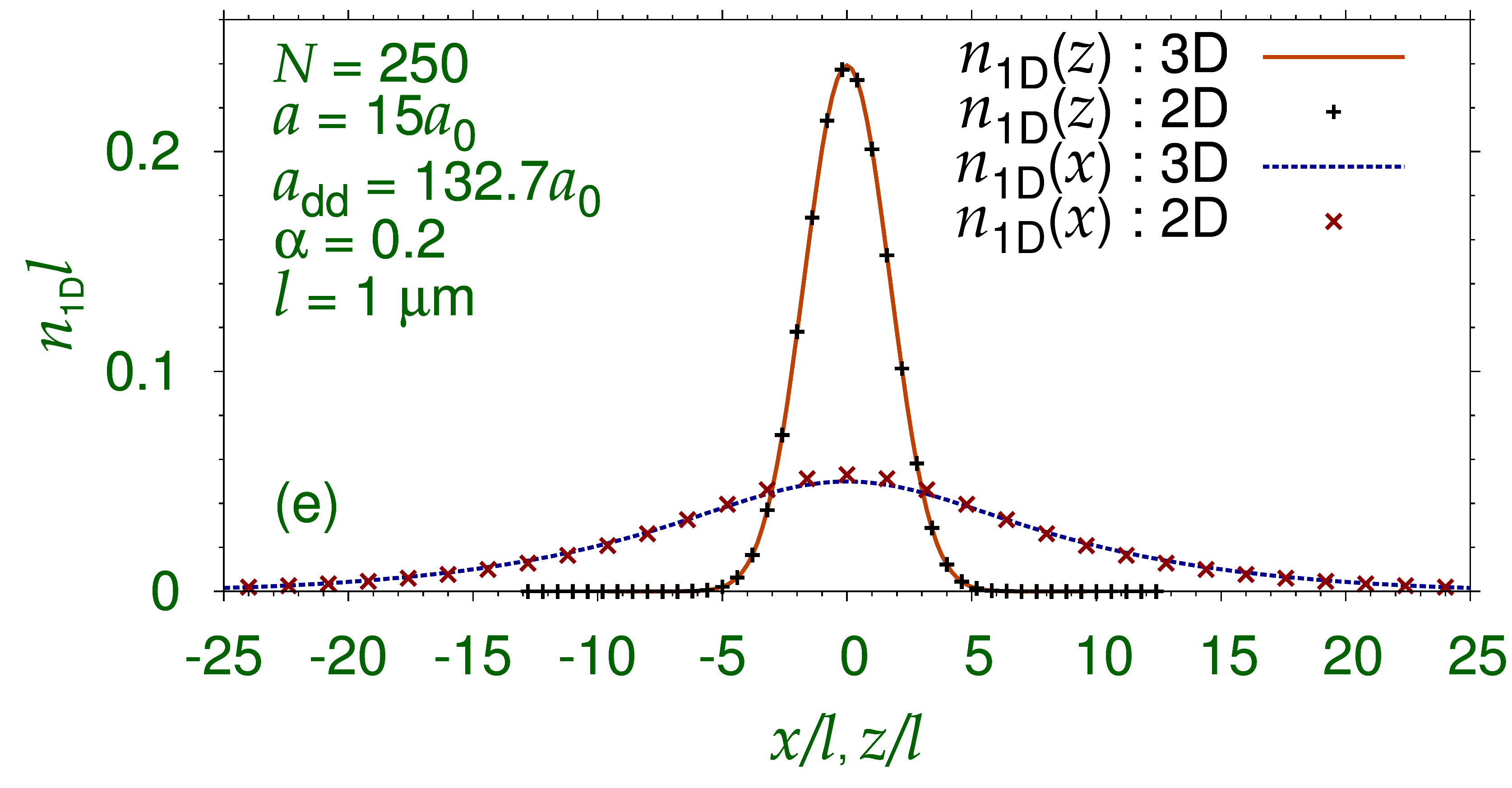}
\includegraphics[width=.49\linewidth,clip]{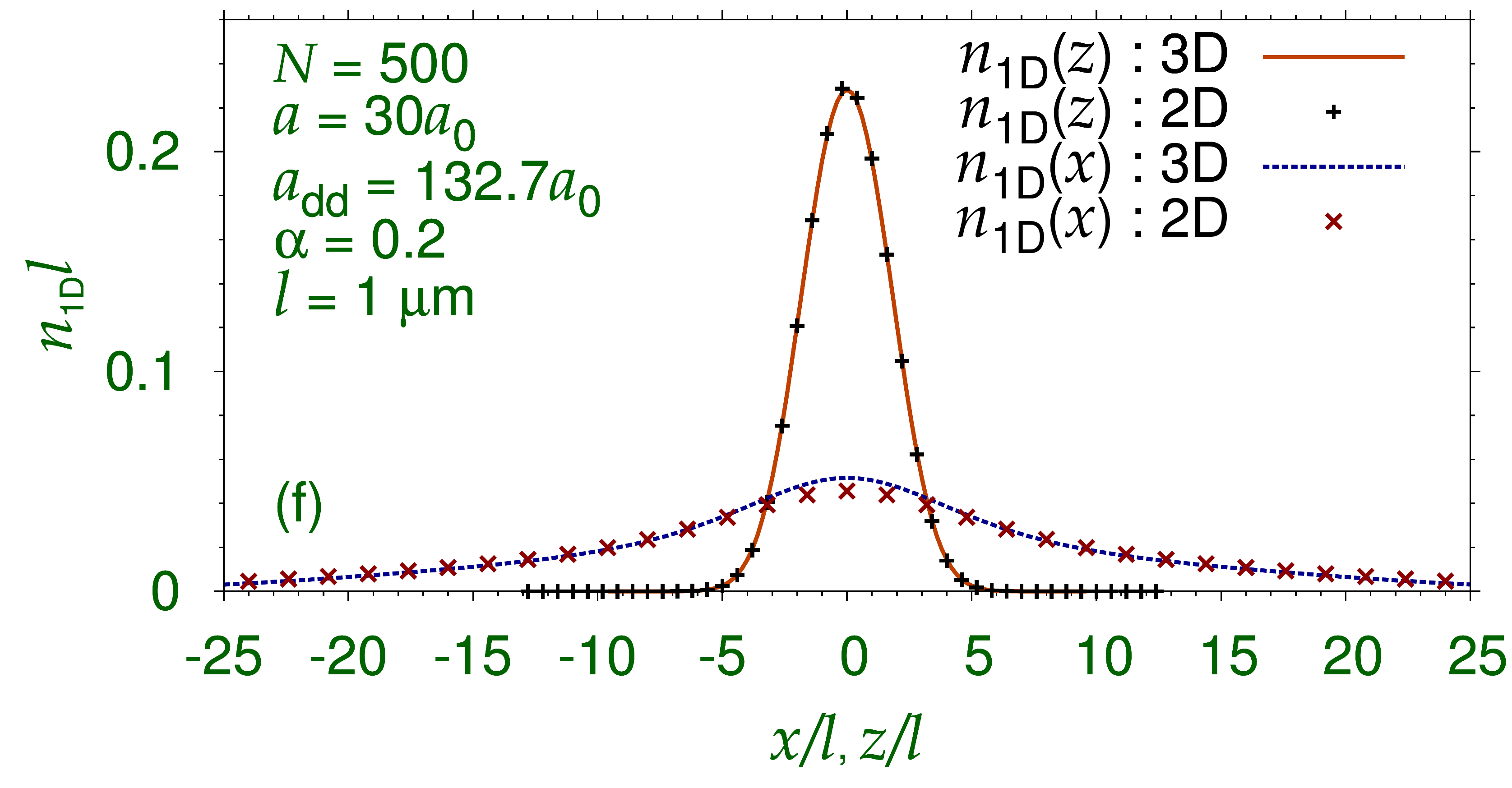}

\caption{ (Color online) Linear densities $n_{1D}(x)$ and $n_{1D}(z)$ for the six solitons 
illustrated in Figs. \ref{fig2} (a), (b), (c), (d), (e) and (f),
as obtained from the 2D and 3D models (\ref{eq3}) and (\ref{gpe2dxz}), in plots (a), (b), (c), (d),
(e), and (f), respectively.}\label{fig3}
\end{center}

\end{figure}

{
Next we study the appropriateness of the reduced 2D model (\ref{gpe2dxz}) for these solitons.} To this end we solve Eq. (\ref{gpe2dxz}) with the same parameters as employed in Figs.
\ref{fig2}.  To compare the results of the 3D and 2D models, we plot the  1D densities, along $x$ and $z$ directions obtained from the 3D and 2D models, defined, for example, by, 
\begin{align}\label{1dx}
&n_{1D}(x) =\int_{-\infty}
^{\infty}dy  \int_{-\infty}
^{\infty}dz |\phi({\bf r})|^2\\   
&n_{1D}(x) =  \int_{-\infty}
^{\infty}dz |\phi_{2D}({\boldsymbol \rho})|^2,
\end{align} 
respectively.   
The linear density $n_{1D}(z)$ is defined similarly. 
In Figs. \ref{fig3} (a) $-$ (f) we plot the densities $n_{1D}(x)$ and $n_{1D}(z)$
for the six solitons illustrated in Figs.  \ref{fig2} (a) $-$ (f), respectively, as calculated using the 3D model (\ref{eq3}) and the 2D model (\ref{gpe2dxz}).  {The agreement between the two sets of densities is  very satisfactory. However, the solitons with larger $\alpha$ (=0.2)
are very long along the $x$ direction and their computation is tedious. Nevertheless, they are robust and would be of experimental interest. This is why for further studies of statics and dynamics we consider only $\alpha=0.05$ in the following. }

\begin{figure}[!t]

\begin{center}
\includegraphics[width=\linewidth]{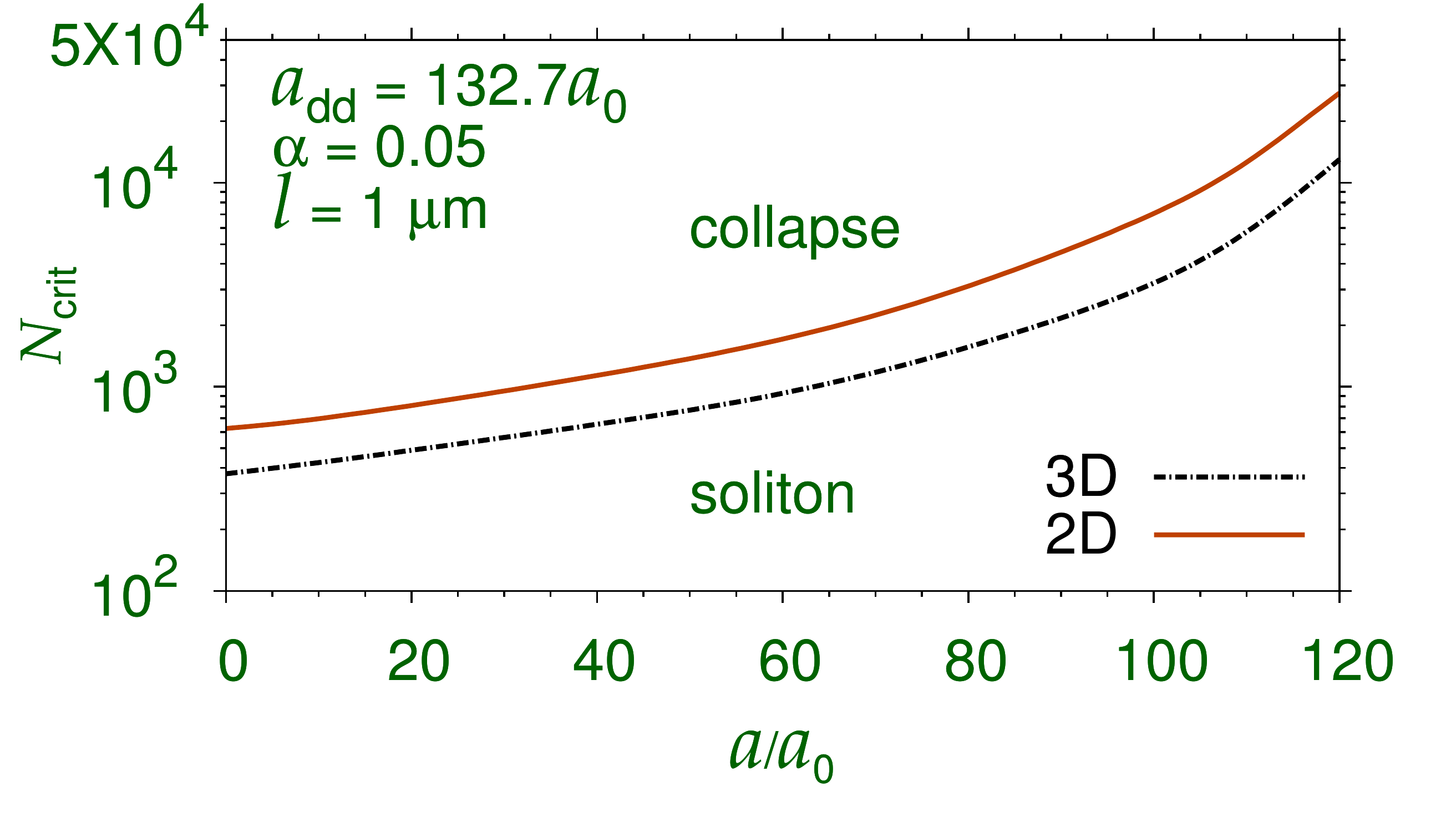}

\caption{ (Color online)  
 Critical number of atoms $N_{\mathrm{crit}} $   for soliton formation from a numerical solution 
of the 3D and 2D models (\ref{eq3}) and (\ref{gpe2dxz}). 
}\label{fig4} \end{center}

\end{figure}

We find that for the  interaction parameters of $^{164}$Dy atoms the  bright soliton is stable up to a critical number $N_{\mathrm{crit}}$ of atoms, beyond which it collapses \cite{jbohn}. We calculate this critical number using the 3D and 2D models and show the result
in Fig. \ref{fig4}.    
A stable soliton is possible for
 $a \lesssim a_{\mathrm {dd}}$ and for a number of atoms below this critical number \cite{1D}. The critical number of atoms 
increases with the increase of contact repulsion.  
An increase of contact repulsion reduces the collapse instability of solitons 
bound by long-range dipolar attraction.
In the region marked soliton in Fig. \ref{fig4} there is a balance between attraction and repulsion and a stable soliton can be formed. In the collapse region, the soliton collapses 
due to an excess of dipolar attraction. 
  The bright solitons are unconditionally stable and 
last for ever in real-time propagation without any visible change of shape.  
In order to have  
robust solitons one should have $a_{\mathrm{dd}} > a > 0$ 
corresponding to a dominating dipolar attraction over a sizable contact repulsion.

\begin{figure}[!t]

\begin{center}
\includegraphics[width=\linewidth]{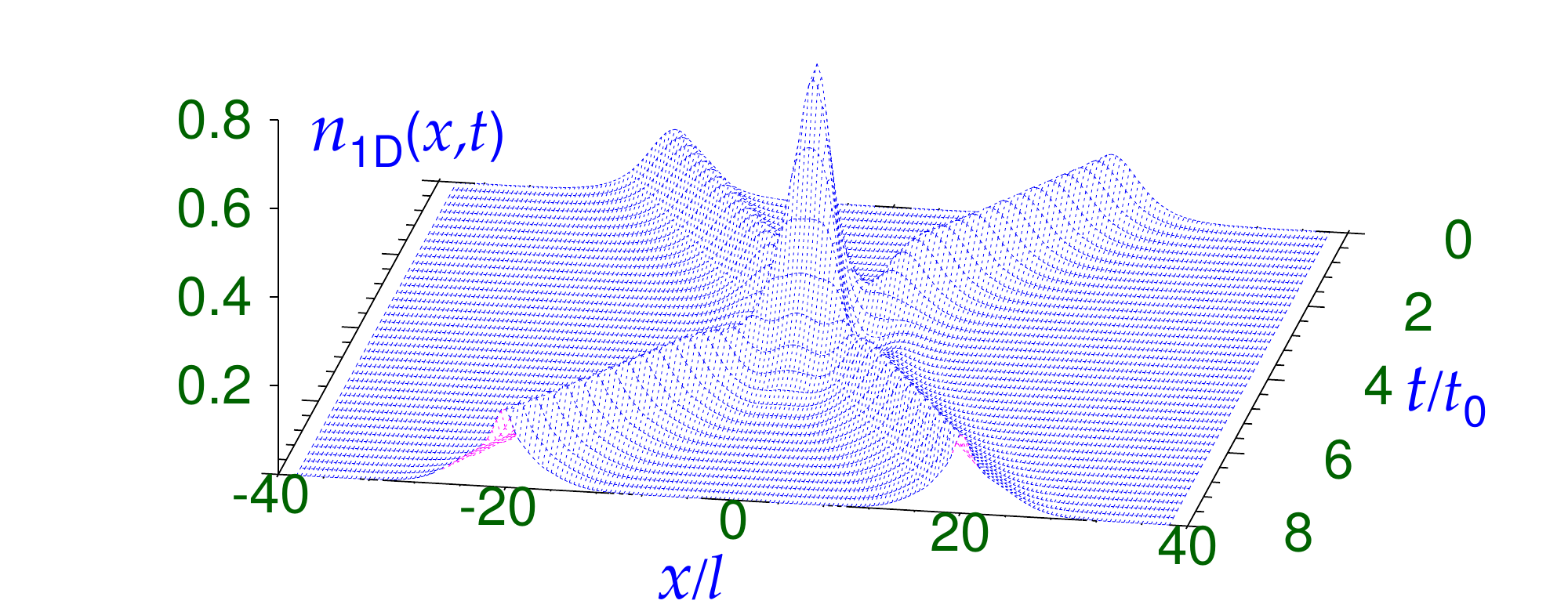}

\caption{ (Color online)  
  Elastic collision dynamics of two solitons of Fig. \ref{fig2} (c) from a solution of the  
3D model (\ref{eq3}) of two identical solitons of Fig. \ref{fig2} (c) in opposite directions
via a plot of linear density $n_{1D}(x,t)$ versus $x$ and $t$.
}\label{fig5} \end{center}

\end{figure}

\begin{figure}[!b]

\begin{center}
\includegraphics[width=\linewidth,clip]{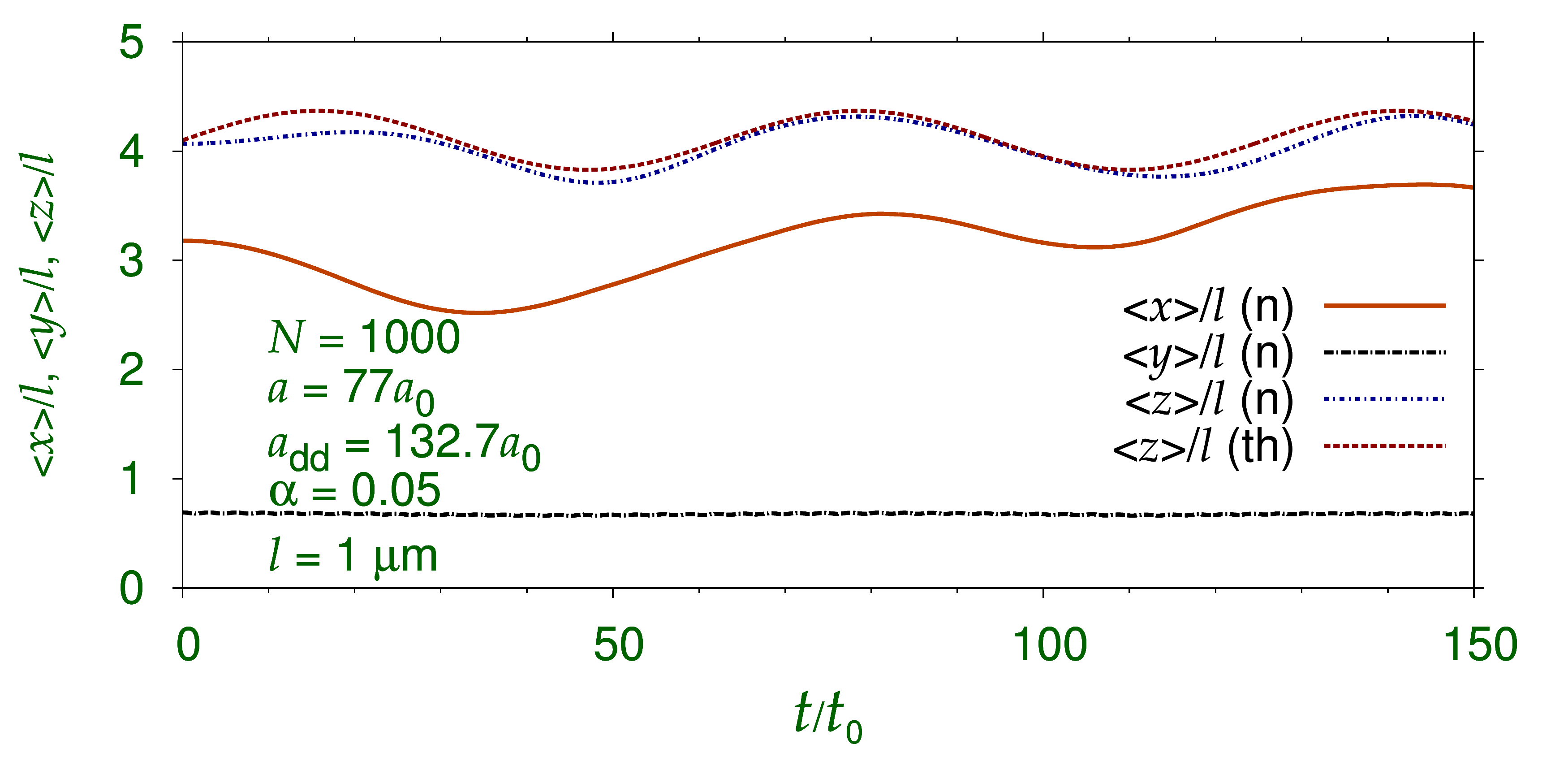}

\caption{(Color online) Numerical (n) result for oscillation dynamics of the soliton of Fig. 
\ref{fig2} (c)
 after multiplying both nonlinearities $g$ and $g_{\mathrm{dd}}$ by 1.2 at $t=0$. 
The theoretical (th) result for oscillation of $\langle z(t)-z(0) \rangle/l \sim \sin(2\alpha t/t_0)$
is also shown for a comparison.
}\label{fig6} \end{center}

\end{figure}

To demonstrate further the robustness of the solitons we consider a head-on collision between two solitons moving along the $x$ axis in opposite directions. We consider the collision between two identical bright solitons of Fig. \ref{fig2} (c) each of 1000 $^{164}$Dy atoms with $a=77a_0 $. 
 and $l=1$ $\mu$m.
The constant velocity of about 1.9 mm/s of each   of the colliding solitons was achieved by phase imprinting the wave function profiles, obtained from an imaginary-time simulation in 3D,
with the  factors 
$\exp(\pm \mathrm{i} 7.5 x)$.
  In Fig. \ref{fig5} we plot the integrated 1D density $n_{1D}(x)$
for the collision dynamics 
obtained from 
Eq. (\ref{1dx}) after a real-time numerical simulation in 3D.
  After the collision, the solitons emerge  without a visible change of shape demonstrating the solitonic nature.  {  However, at   lower incident velocities the collision becomes inelastic and   distortion in the shape of the solitons is noted. }

As a further test of stability of the solitons, we demonstrate their sustained oscillation under a sizable perturbation.  
We perform a real-time simulation of the 3D profile of the soliton of Fig. \ref{fig2} (c) upon multiplying  both the nonlinearities $g$ and 
$g_{\mathrm{dd}}$ in Eq. (\ref{eq3})  by the factor 1.2. The soliton executes breathing oscillation 
as illustrated in Fig. \ref{fig6} where we plot the root-mean-square  sizes 
$\langle x\rangle ,\langle y\rangle,\langle z \rangle$, respectively, 
in directions $x, y$, and $z$ versus time.  We also plot in this  case the  theoretical result for 
oscillation with frequency $2\alpha$ \cite{str}
along $z$ direction
and this compares vary favorably with the numerical result.

\section{Conclusion}

We demonstrated  the possibility of creating stable bright solitons in dipolar BEC
mobile in a direction ($x$) perpendicular to the polarization direction ($z$) using the full 
mean-field GP equation in 3D and a reduced 2D GP equation in the plane $x-z$. 
{ The reduced 2D model is found to yield results in excellent agreement with the full 3D model.}
The head on collision between two identical solitons with relative velocity of about 
4 mm/s is demonstrated to be quasi elastic with the solitons passing through each other with practically no deformation.   The solitons are found to execute sustained breathing oscillation 
upon a small perturbation. 
 The numerical simulation was done by explicitly solving 
the  GP equation with realistic values of contact and  dipolar interactions of $^{164}$Dy    atoms.  The results and conclusions  of the present paper can be  tested in experiments with present-day know-how and 
technology  and should lead to interesting future investigations.

We thank  
FAPESP  and  CNPq (Brazil)  for partial support.

\end{document}